\documentclass{rnaastex}

\begin{document}

\title{Galaxy Nurseries: Crowdsourced analysis of slitless spectroscopic data}

\correspondingauthor{Hugh Dickinson}
\email{hdickins@umn.edu}

\author[0000-0003-0475-008X]{Hugh Dickinson}
\affiliation{School of Physics and Astronomy, Tate Laboratory of Physics, 116 Church St. SE, Minneapolis, MN 55455, United States of America}

\author[0000-0002-9136-8876]{Claudia Scarlata}
\affiliation{School of Physics and Astronomy, Tate Laboratory of Physics, 116 Church St. SE, Minneapolis, MN 55455, United States of America}

\author[0000-0002-1067-8558]{Lucy Fortson}
\affiliation{School of Physics and Astronomy, Tate Laboratory of Physics, 116 Church St. SE, Minneapolis, MN 55455, United States of America}

\author[0000-0002-9921-9218]{Micaela Bagley}
\affiliation{School of Physics and Astronomy, Tate Laboratory of Physics, 116 Church St. SE, Minneapolis, MN 55455, United States of America}

\author[0000-0001-7166-6035]{Vihang Mehta}
\affiliation{School of Physics and Astronomy, Tate Laboratory of Physics, 116 Church St. SE, Minneapolis, MN 55455, United States of America}

\author{John Phillips}
\affiliation{School of Physics and Astronomy, Tate Laboratory of Physics, 116 Church St. SE, Minneapolis, MN 55455, United States of America}

\author[0000-0003-0556-2929]{Ivano Baronchelli}
\affiliation{IPAC, Mail Code 314-6, Caltech, 1200 E. California Blvd., Pasadena, CA 91125, United States of America}

\author[0000-0002-7928-416X]{Sophia Dai}
\affiliation{National Astronomical Observatories Chinese Academy of Sciences: Chaoyang-qu, Beijing, China}

\author[0000-0001-6145-5090]{Nimish Hathi}
\affiliation{Space Telescope Science Institute, 3700 San Martin Drive, Baltimore, MD 21218, United States of America}

\author[0000-0002-6586-4446]{Alaina Henry}
\affiliation{Space Telescope Science Institute, 3700 San Martin Drive, Baltimore, MD 21218, United States of America}

\author[0000-0001-6919-1237]{Matthew Malkan}
\affiliation{Physics and Astronomy Department, University of California, Los Angeles, CA 90095-1547, USA}

\author[0000-0002-9946-4731]{Marc Rafelski}
\affiliation{Space Telescope Science Institute, 3700 San Martin Drive, Baltimore, MD 21218, United States of America}
\affiliation{Department of Physics \& Astronomy, Johns Hopkins University, Baltimore, MD 21218, USA}

\author[0000-0002-7064-5424]{Harry Teplitz}
\affiliation{IPAC, Mail Code 314-6, Caltech, 1200 E. California Blvd., Pasadena, CA 91125, United States of America}

\author[0000-0001-8600-7008]{Anita Zanella}
\affiliation{European Southern Observatory, Karl Schwarzschild Stra\ss e 2, 85748 Garching, Germany}

\author[0000-0001-5578-359X]{Chris Lintott}
\affiliation{Oxford Astrophysics, University of Oxford, Denys Wilkinson Building, Keble Road, Oxford, OX1 3RH, UK}


\section{The Galaxy Nurseries Project}\label{sec:theGNProject}

The \textit{Galaxy Nurseries}\footnote{\url{www.zooniverse.org/projects/hughdickinson/galaxy-nurseries}} project was designed to enable crowdsourced analysis of slitless spectroscopic data by volunteers using the \textit{Zooniverse}\footnote{\url{www.zooniverse.org}} online interface. The dataset was obtained by the WFC3 Infrared Spectroscopic Parallel (WISP) Survey collaboration \citep{2010ApJ...723..104A} and comprises NIR grism (G102 and G141) and direct imaging of 432 fields. The scientific goals of WISP Survey require reliable identification of emission lines \citep[e.g.][]{0004-637X-785-2-153,0004-637X-789-2-96}. Spectral contamination by overlapping signals from multiple sources or diffraction orders as well as residual artifacts of the data reduction can significantly complicate the automatic detection and identification of emission lines.  Visual verification of automatically detected features has proved essential to obtain a pure sample of emission lines.

In \textit{Galaxy Nurseries} verification of putative emission lines was delegated to citizen-scientist volunteers. Data were presented as ``subject images'' using the fixed format illustrated in the top panels of Figure \ref{fig:multipanel_figure}. For each target, volunteers were provided with the two-dimensional spectrum (B), the corresponding one-dimensional extraction (A) and its direct image (C). Volunteers were instructed to evaluate only the feature identified by the green crosshairs and to decide whether it was a genuine emission line or more likely an artifact, providing a Boolean Valued (i.e. ``Real'' or ``Spurious'') label. Following its launch, \textit{Galaxy Nurseries} was completed in only 40 days, gathering 414,360 classifications from 3003 volunteers for 27,333 putative emission lines. At least 15 classifications were obtained for each subject image. For reference, it took approximately 4.5 months for the full sample of lines to be visually inspected by two members of the WISP Survey Science Team (WSST).

\section{Results}\label{sec:results}
\begin{figure}[h!]
\begin{center}
\includegraphics[height=0.3\textwidth]{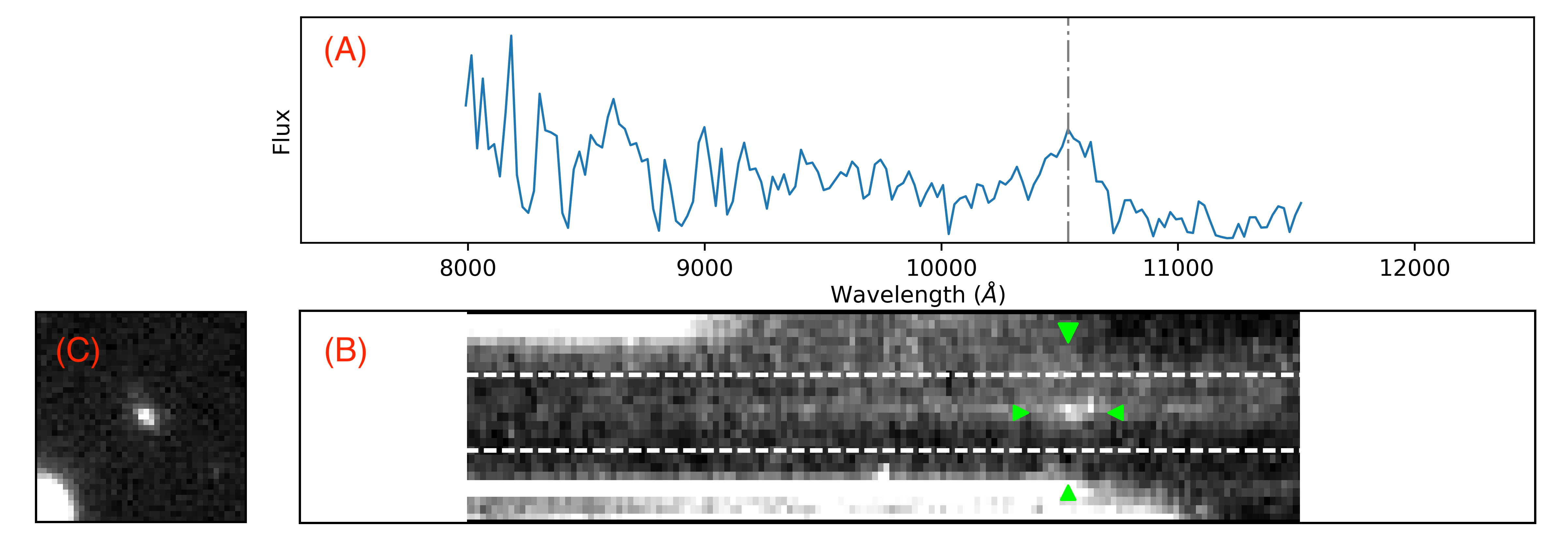}
\includegraphics[height=0.3\textwidth]{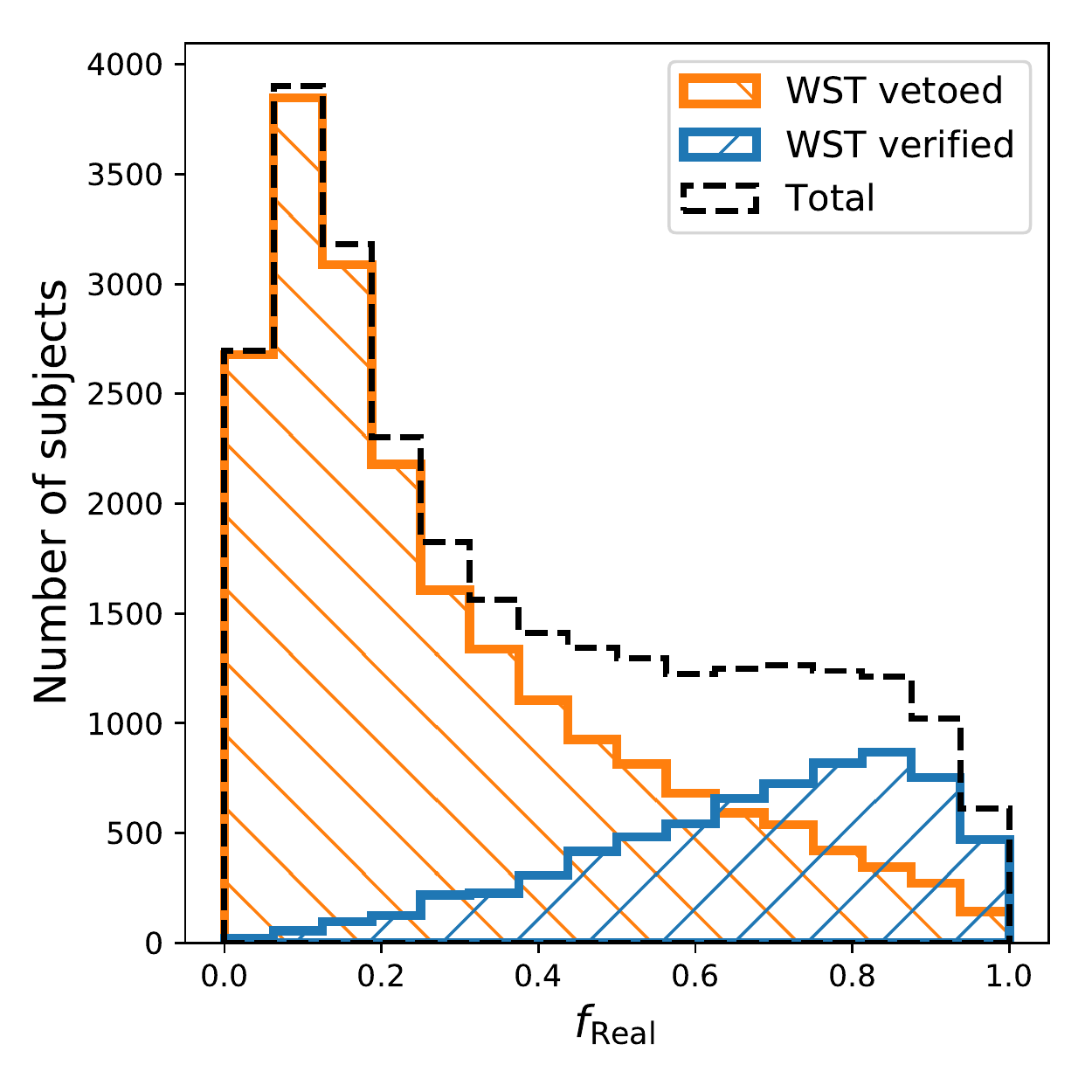}
\includegraphics[height=0.3\textwidth]{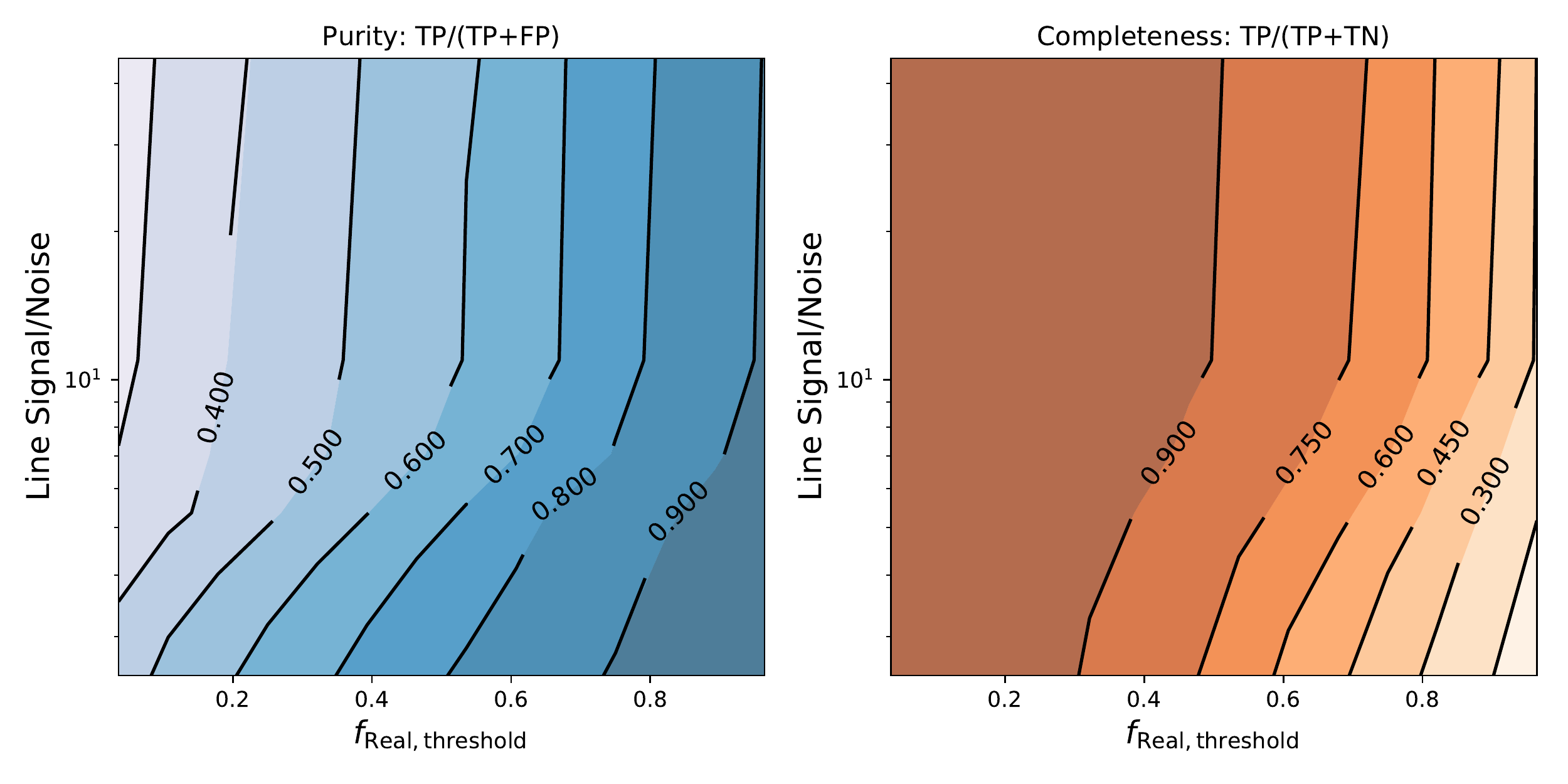}

\caption{\textit{Top:} Typical fixed-format subject image used by volunteers to classify putative emission lines as real or spurious. \textit{Lower left:} The distribution of $f_{\mathrm{Real, threshold}}$ for the full sample (\textit{black, dashed}), the sub-sample of WSST-verified lines (\textit{blue}) and the sub-sample of WSST-vetoed lines (\textit{orange}). Sample purity ($P_{S}=TP/(TP+FP)$, \textit{lower centre}) and completeness ($C_{S}=TP/(TP+TN)$, \textit{lower right}) realised using volunteer classifications as a function of threshold subject vote fraction $f_{\mathrm{Real, threshold}}$ and line signal-to-noise ratio.
\label{fig:multipanel_figure}}
\end{center}
\end{figure}

Volunteer responses for each subject were aggregated to compute $f_{\mathrm{Real}}$, the fraction of volunteers who classified the corresponding emission line as ``Real''. To evaluate the accuracy of volunteer classifications, their aggregated responses were compared with independent assessments provided by members of the WSST\footnote{Full results of the WSST evaluation will be presented in Bagley et al. 2019 (in prep.)}. The distribution of $f_{\mathrm{Real}}$ for all subjects is shown in the \textit{bottom-left} panel of Figure~\ref{fig:multipanel_figure} and is subdivided to identify emission lines that were verified (blue) and vetoed (orange) by the WSST. The relative small number of lines that were validated by the WSST underlines the need for visual inspection of the full sample. Overall, there is a broad agreement between the WSST and volunteers classifications.
The distribution of $f_{\mathrm{Real}}$ peaks at 0.85 for the sample of verified lines, indicating that the majority of volunteers agree on the reality of these lines. Similarly, the distribution of $f_{\mathrm{Real}}$ peaks at 0.1 for the sample of vetoed lines.
It is clear, however, that the agreement is not perfect: there is a large number of lines which were vetoed by the WSST and would be classified as real by most of the volunteers (about one third of the objects with $f_{\mathrm{Real}}>0.7$ have been vetoed by the WSST). The opposite is not true: only a minority of lines with low $f_{\mathrm{Real}}$ were in fact real according to the WSST. This result suggests that, with a minimal level of training, volunteers are successful in securely identifying artifacts in the data.

The result is reinforced by a more quantitative analysis that accounts also for the emission line signal-to-noise ratio ($S/N$).  The remaining \textit{middle} panels of Figure \ref{fig:multipanel_figure}  illustrate how the purity ($P_{S}=TP/(TP+FP)$)\footnote{The labels T(rue)/F(alse) are assigned to lines verified/vetoed by the WSST, while lines for which $f_{\mathrm{Real}}$ greater/smaller than $f_{\mathrm{Real, threshold}}$ are labeled as P(ositive)/N(egative).} and completeness ($C_{S}=TP/(TP+TN)$) change as a function of $f_{\mathrm{Real, threshold}}$ and $S/N$.
These results show that the degree of categorical separation provided by volunteer classifications may be adequate in certain situations. If purity is paramount, then for $S/N\gtrsim5$, choosing $f_{\mathrm{Real, threshold}}\sim 0.7$ yields a sample purity $P_{S}\gtrsim0.8$. The corresponding sample completeness is $C_{S}\sim0.6$. Conversely, a more complete ($C_{S}\sim0.9$) albeit impure ($P_{S}\sim0.6$) sample is feasible for $S/N\gtrsim10$ if $f_{\mathrm{Real, threshold}}\sim 0.5$.

The results of \textit{Galaxy Nurseries} demonstrate the feasibility of identifying genuine emission lines in slitless spectra by citizen scientists. We recognise that robust scientific analyses typically require samples with higher purity and completeness than raw volunteer classifications provide. Nonetheless, choosing optimal values for $f_{\mathrm{Real, threshold}}$ allows a large fraction of spurious lines to be vetoed, substantially reducing the timescale for subsequent professional analysis of the remaining potential lines.

\acknowledgments
HD, CS, and LF acknowledge partial support from the US National Science Foundation Grant AST-1413610.
Support for HST Programs GO-11696, 12283, 12568, 12902, 13517, 13352, and 14178 was provided by NASA through grants from the Space Telescope Science Institute, which is operated by the Association of Universities for Research in Astronomy, Inc., under NASA contract NAS5-26555.
This publication uses data generated via the Zooniverse.org platform, development of which is funded by generous support, including a Global Impact Award from Google, and by a grant from the Alfred P. Sloan Foundation.

\bibliographystyle{aasjournal}
\bibliography{GalaxyNurseries}

\end{document}